\documentclass[12pt]{article}
\usepackage[mathscr]{eucal}
\usepackage{amssymb,latexsym}
\usepackage{verbatim}
\usepackage{amsmath}
\usepackage{amsthm}
\usepackage{enumerate}
\usepackage{authblk}
\usepackage{color}
\usepackage{url}

\numberwithin{equation}{section}

\newtheorem{thm}{Theorem}[section]
\newtheorem{lem}[thm]{Lemma}

\newcommand\calL{{\mathcal{L}}}

\renewcommand\l{\lambda}

\renewcommand\S{\Sigma}

\renewcommand\d{\partial}

\newcommand\f{\phi}

\newcommand\D{\nabla}
\newcommand\e{\epsilon}

\renewcommand\div{{\rm div}}

\newcommand\<{\langle}
\renewcommand\>{\rangle}

\renewcommand\l{\lambda}
\newcommand\g{\gamma}

\renewcommand\t{\tau}
\renewcommand\th{\theta}

\newcommand\beq{\begin{equation}}
\newcommand\eeq{\end{equation}}
\newcommand\ben{\begin{enumerate}}
\newcommand\een{\end{enumerate}}
\newcommand\bit{\begin{itemize}}
\newcommand\eit{\end{itemize}}

\DeclareMathOperator{\diver}{div}
\DeclareMathOperator{\Real}{Re}

\renewcommand{\div}{\diver}
\renewcommand{\Re}{\Real}

\newcommand{\tr}{\mathrm{tr}\,}

\newcounter{mnotecount}

\title{Rigidity of outermost MOTS - the initial data version}

\author{Gregory J. Galloway\thanks{Research partially supported by NSF grant DMS-1710808.}}

\affil{Department of Mathematics
\\University of Miami}

\begin{document}
\date{}
\maketitle  
\vspace{.2in}

\begin{abstract}  
In \cite{Grigid}, a rigidity result was obtained for outermost marginally outer trapped surfaces (MOTSs) that do not admit metrics of positive scalar curvature. This allowed one to treat  the  ``borderline case"  in the author's work with R. Schoen concerning the topology of higher dimensional black holes \cite{GS}.   The proof of this rigidity result involved bending the initial data manifold in the vicinity of the MOTS within the ambient spacetime.  In this note we show how to circumvent this step, and thereby obtain a pure initial data version of this rigidity result and its consequence concerning the topology of black holes. 

\end{abstract}


\section{Introduction}
\label{intro}

The aim of this note is to obtain pure initial data versions  of the main results in \cite{Grigid} concerning the topology and rigidity of marginally outer trapped surfaces.   This settles problem (18) in the open problem section in \cite{BAMS}.  In order to describe our results and put them in context, we begin with some basic definitions and  background.

Let $(M,g)$ be an $n+1$, $n \ge 3$, dimensional spacetime (time oriented Lorentzian manifold).   By an initial data set in $(M,g)$ we mean a triple $(V, h, K)$, where $V$ is a smooth spacelike hypersurface, $h$ is its induced   (Riemannian) metric and $K$ is its second fundamental form.  To set sign conventions, we have $K(X,Y) = g(\nabla_Xu,Y)$, where $X,Y \in T_pV$, $\D$ is the Levi-Civita connection of $M$, and $u$ is the future directed timelike unit vector field to $V$. 

Recall that  the {\it spacetime} dominant energy condition is the requirement,
\beq\label{dec}
G(X,Y) \ge 0  \quad \text{for all future directed causal vectors $X,Y \in TM$},
\eeq
where $G =  Ric_M - \frac12 R_M g$ is the Einstein tensor.  Given an initial data set $(V,h,K)$, the spacetime dominant energy condition implies
\beq\label{iddec}
\mu \ge |J| \quad \text{along $V$} \,,
\eeq
where the scalar $\mu = G(u,u)$  is the {\it local energy density}  and  the one-form $J = G(u,\cdot)$ is the {\it local momentum density} along $V$.  It is a basic fact that $\mu$ and $J$ can be expressed solely in terms of initial data. When referring to an initial data set, the inequality \eqref{iddec} is what is meant by the dominant energy condition.

We now recall the key concept of a marginally outer trapped surface.
Consider an initial data set $(V, h, K)$ in a spacetime $(M,g)$.  Let $\S$ be a closed (compact without boundary)
two-sided hypersurface in $V$. Then $\S$ admits a smooth unit normal field
$\nu$ in $V$, unique up to sign.  By convention, refer to such a choice as outward pointing. Let $u$ be the future pointing timelike unit vector field orthogonal to $M$.
Then $l_+ = u+\nu$ (resp. $l_- =  u - \nu$) is a future directed outward (resp., future directed inward) pointing null normal vector field along $\S$.

Associated to $l_+$ and $l_-$,  are the two {\it null second fundamental forms},  $\chi_+$ and 
$\chi_-$, respectively, defined as, 
$\chi_{\pm} : T_p\S \times T_p\S \to \mathbb R $,  $\chi_{\pm}(X,Y) = g(\D_Xl_{\pm}, Y)$.
The {\it null expansion scalars} (or {\it null mean curvatures})  $\th_{\pm}$ of $\S$   are obtained by tracing 
$\chi_{\pm}$,
\beq
\theta_\pm=\tr \chi_\pm=\tr_\Sigma K\pm H \,,
\eeq 
where, in the latter expressions, which depend only on initial data,  $H$ is the mean curvature of $\Sigma$ in $V$ and ${\rm tr}_{\S} K$ is the partial trace of $K$ along $\S$.
Physically, $\th_+$ (resp., $\th_-$) measures the divergence
of the  outgoing (resp., ingoing) light rays emanating
from $\S$. 

In regions of spacetime where the
gravitational field is strong, one may have both $\th_- < 0$
and $\th_+ < 0$, in which case $\S$ is called a {\it trapped
surface}, the important concept introduced by Penrose. 
Focusing attention on the outward null normal only, we say that
$\S$ is an {\it outer trapped surface}  if $\th_+ < 0$.  Finally, we say that $\S$ is a {\it marginally
outer trapped surface} (MOTS) if $\th_+$ vanishes identically.   
MOTSs arise naturally in a number of situations.  For example,  cross sections of the event horizon in stationary black holes spacetimes, such as the Kerr solution, are MOTSs.  MOTSs may also occur as the boundary of the `trapped region' (cf. \cite{AEM}, and references therin).  We note that in the 
time-symmetric case ($K = 0$), a MOTS is a minimal surface in $V$.

A basic step in the proof of the ``no hair theorem" (i.e., the uniqueness of the Kerr solution) is Hawking's theorem on the topology of black holes~\cite{HE}, which asserts that compact cross-sections of the event horizon in
$3+1$-dimensional asymptotically flat
stationary black hole space-times obeying the dominant energy
condition are topologically 2-spheres.   The  discovery of
Emparan and Reall~\cite{ER} of a $4+1$ dimensional asymptotically flat
stationary vacuum black hole space-time with horizon topology $S^1
\times S^2$, the so-called ``black ring", showed that black hole uniqueness fails in higher dimensions and, moreover, that horizon topology need not be spherical. This naturally led to the
question as to what  horizon topologies are allowed in higher
dimensional black hole space-times. This question was addressed
in a  paper  with R. Schoen \cite{GS}, in which a generalization of Hawking's 
black hole topology theorem was obtained.  

Let $\S$ be a MOTS in an initial data set $(V,h,K)$,  and suppose 
$\S$ separates $V$ into an ``inside" and an ``outside" (the side into which $\nu$ points).
We shall say that $\S$ is {\it outermost} in $V$ if there are no outer trapped ($\th_+  < 0$)  or marginally outer trapped ($\th_+= 0$) surfaces outside of and homologous to $\S$.    Outermost MOTS are necessarily {\it stable}.
(The important concept of the stability of MOTS \cite{AMS1,AMS2} is reviewed in the next section.)  In \cite{GS} the author  and Schoen proved the following.

\begin{thm}
\label{bhtopo}
Let $(V,h,K)$ be an $n$-dimensional initial data set, $n \ge 3$, satisfying the dominant energy condition  (DEC),  
$\mu \ge |J|$.  If $\S$ is a stable MOTS  in $V$ (in particular if $\S$ is outermost)
then, apart from certain exceptional circumstances, $\S$
must admit a metric of positive scalar curvature. 
\end{thm}

 The `exceptional circumstances' are ruled out if, for example, the DEC holds strictly at some point of  $\S$ or $\S$ is not Ricci flat.   Apart from such exceptional circumstances, $\S$
is admits a metric of positive scalar curvature, which implies many well known restrictions on the
topology; see \cite{Gbhvol} for a discussion.  In particular, in the case 
${\rm dim}\, M =4+1$, so that
${\dim}\, \S = 3$ (and assuming orientablity),  $\S$ must be diffeomorphic to either a spherical space (quotient of a $3$-sphere) or to $S^1\times S^2$, or to a connected sum of these two types.  

One drawback of Theorem \ref{bhtopo} is that it allows  certain possibilities that one would like to rule out.  For example, it does not rule out the possibility of a vacuum black hole spacetime with toroidal topology.  However, in \cite{Grigid},  we were able
to eliminate these exceptional cases for outermost MOTSs provided the initial data set can be embedded into a spacetime obeying the spacetime DEC \eqref{dec}.

\begin{thm}[\cite{Grigid}]\label{posyam2}
Let $(V^n,h,K)$, $n \ge 3$, be an initial data set in a spacetime obeying the DEC. If 
$\S^{n-1}$ is an outermost  MOTS in $(V^n,h,K)$ then  $\S$ admits a metric of positive scalar curvature. 
\end{thm}

Thus, in particular, there can be no stationary vacuum black hole spacetime  with toroidal horizon topology.
Theorem~\ref{posyam2} is an immediate consequence of the following rigidity result.

\begin{thm}[\cite{Grigid}]\label{rigid}
Let $(V^n,h,K)$, $n \ge 3$, be an initial data set in a spacetime obeying the DEC. 
 Suppose  $\S$ is a separating MOTS in $V$  such that there are
no outer trapped surfaces ($\th_+ < 0$) outside of, and homologous, to $\S$.  If $\S$ does not admit a metric of positive scalar curvature, then there exists
an outer half-neighborhood $U \approx [0,\e) \times \S$ of $\S$ in $V$ such that each slice
$\S_t = \{t\} \times \S$, $t \in [0,\e)$ is a MOTS. 
In fact, each $\S_t$  has vanishing null second fundamental form, with respect to the outward null normal, and is Ricci flat.
\end{thm}

An unsatisfactory feature of Theorems \ref{posyam2} and \ref{rigid}, from both a conceptual and practical point of view, is that they are not pure initial data results: The proof of Theorem \ref{rigid} in \cite{Grigid} requires the DEC   \eqref{dec} to hold in a spacetime neighborhood of $\S$.    However, many fundamental results in general relativity, such as the positive mass theorem, are statements  about initial data sets; no assumptions about the evolution of the data are required.   From this point of view, it would be desirable to obtain a pure initial data version of Theorem \ref{rigid}, one that only requires the initial data version \eqref{iddec} of the DEC.  Such a version is presented in Section 3.  Some preliminary results are presented in Section 2.

\section{Preliminaries}\label{prelim}

Let $(\S,\g)$ be a compact Riemannian manifold.  We will be considering operators $\calL : C^{\infty}(\S) \to C^{\infty}(\S)$ of the form
\begin{align}\label{op}
\mathcal{L}(\f) = - \triangle \f   + 2\<X,\D \f\> + (\mathcal{Q} + \div X - |X|^2) \f  \,,
\end{align}
where $\mathcal{Q} \in C^{\infty}(\S)$, $X$ is a smooth vector field on $\S$ and 
$\<\,,\,\> = \g$.  

Although the operator $\calL$ is not self-adjoint in general,  it nevertheless has the following properties (see \cite{AMS2}).

\begin{lem}\label{prin}
The following holds for the operator $\calL$.
\begin{enumerate}
\item There is a real eigenvalue $\lambda_1=\lambda_1(\calL)$, called the {\em principal eigenvalue} of $\calL$, such that for any other eigenvalue $\mu$, $\Re(\mu)\ge\lambda_1$. The associated eigenfunction $\phi$, $L\phi=\lambda_1\phi$, is unique up to a multiplicative constant, and can be chosen to be strictly positive.
\item $\lambda_1\ge0$ (resp., $\lambda_1>0$) if and only if there exists $\psi\in C^\infty(\Sigma)$, $\psi>0$, such that $\calL(\psi) \ge 0$ (resp., $\calL(\psi) > 0$).  
\end{enumerate}
\end{lem}

\medskip

The following is proved in \cite{Grigid} (based on the main argument in \cite{GS}).

\begin{lem}\label{scalar} 
Consider the operator $\calL$ 
such that,
\begin{align}\label{Q}
\mathcal{Q} = \frac12 S - P \,,
\end{align}
where $S$ is the scalar curvature of $(\S,\g)$ and $P \ge 0$. 
If $\l_1(\calL) \ge 0$ then $\S$ admits a metric of positive scalar curvature,
unless $\l_1(\calL) = 0$, $P \equiv 0$ and $(\S,\g)$ is Ricci flat.  

\end{lem}

MOTSs admit an important notion of stablilty, as introduced by Andersson, Mars and Simon  \cite{AMS1,AMS2}, which we now recall.  In what follows, to simplify notation, we drop the plus sign, and denote $\theta=\theta_+$, $\chi=\chi_+$, and $l=l_+$.  

Let $\S$ be a MOTS in the initial data set $(V,h,K)$ with outward unit normal $\nu$.  We consider a normal variation of $\S$ in $V$,  i.e.,  a variation 
$t \to \S_t$ of $\S = \S_0$ with variation vector field 
$V = \frac{\d}{\d t}|_{t=0} = \phi\nu$,  $\phi \in C^{\infty}(\S)$.
Let $\th(t)$ denote the null expansion of $\S_t$
with respect to $l_t = u + \nu_t$, where $u$ is the future
directed timelike unit normal to $M$ and $\nu_t$ is the
outer unit normal  to $\S_t$ in $M$.   A computation as in \cite{AMS2} gives,
\beq\label{thder} 
\left . \frac{\d\th}{\d t} \right |_{t=0}   =
L(\f) \;, 
\eeq 
where $L : C^{\infty}(\S) \to C^{\infty}(\S)$ is the operator, 
\beq\label{stabop}
L(\phi)  = -\triangle \phi + 2\<X,\D\phi\>  + \left(Q +{\rm div}\, X - |X|^2 \right)\phi \,,
\eeq 
and where,
\beq\label{Q}
Q =  \frac12 S_{\S} - (\mu + J(\nu)) - \frac12 |\chi|^2\,.
\eeq
Here, $\triangle$, $\D$ and ${\rm
div}$ are the Laplacian, gradient and divergence operators,
respectively, on $\S$, $S_{\S}$ is the scalar curvature of $\S$ with respect to the induced metric 
$\<\,,\,\>$ on $\S$, $X$ is the 
vector field  on $\S$  dual to the one form $K(\nu,\cdot)|_{T\S}$, 
and $\mu$ and $J$ are as in the introduction.

We note that $L$ is of the form \eqref{op}.  In the time-symmetric ($K = 0$)
case, $L$ reduces to the classical stability (or Jacobi) operator of minimal surface theory.  As such,  $L$ is referred to as the MOTS stability operator.   
We say that a MOTS is stable provided $\l_1(L) \ge 0$.  In the minimal surface case
this is equivalent to the second variation of area being nonnegative. Lemma \ref{prin} and \eqref{stabop} imply that a MOTS is stable if and only there is an outward variation $t  \to \S_t$ such that 
$\frac{\d\th}{\d t}|_{t=0} \ge 0$.  

A basic criterion for stability is the following. We say that a separating MOTS $\S$ is {\it weakly outermost} provided there are no outer trapped ($\th < 0$) surfaces outside of, and homologous to, $\S$.  Weakly outermost MOTS are necessarily stable.  Indeed, if $\l_1(L) < 0$,  \eqref{thder}, with $\phi$ a positive eigenfunction ($L(\phi) = \l_1(L) \phi$) implies that $\S$ can be deformed outward to an outer trapped surface.
 
\smallskip
The following was a key element in the proofs of Theorems 1.2 and 3.1 in \cite{Grigid} (see also \cite{GMendes, Mendes}).

\begin{lem}\label{foliation}
Let $\Sigma$ be a  MOTS in an initial data set $(V,h,K)$. If $\lambda_1(L)=0$, where $L$ is the MOTS stability operator, then, up to isometry, there exists a neighborhood $W$ of $\S$ such that:
\ben
\item[(i)] $W = (-t_0,t_0) \times \S$ and $h|_W$ has the orthogonal decomposition,
$$
h|_W = \phi^2 dt^2 + \g_t  \,
$$
where $\phi = \phi(t,x)$ and $\g_t$ is the induced metric on $\S_t = \{t\} \times \S$.
\item[(ii)] The outward null expansion of each $\S_t$ is constant, i.e., $\th = \th(t)$, with respect to  $\ell_t = u + \nu_t$, where $\nu_t = \frac{1}{\phi} \frac{\d}{\d t}$ is the outward unit normal to $\S_t$.
\een
\end{lem}

\section{Main results}

The main aim of this section is to prove the following. 

\begin{thm}\label{rigid2}
Let $(V^n,h,K)$, $n \ge 3$, be an initial data set satisfying the DEC, $\mu \ge |J|$. 
Suppose $\S^{n-1}$ is a weakly outermost MOTS in  $V^n$
that does not admit a metric of positive scalar curvature.  Then there exists
an outer neighborhood $U \approx [0,\e) \times \S$ of $\S$ in $V$ such that each slice
$\S_t = \{t\} \times \S$, $t \in [0,\e)$ is a MOTS.  In fact each  such slice
has vanishing outward null second fundamental form, and is Ricci flat.
\end{thm}

Theorem \ref{rigid2} was proved in \cite{Grigid} under the additional assumption that the mean curvature of $\tau = \tr K$ of $V$ is nonpositive, $\tau \le 0$ (cf. \cite[Theorem 3.1]{Grigid}, whose proof only requires \eqref{iddec}). This assumption was removed in \cite[Theorem 1.2]{Grigid}, assuming that the ambient spacetime satisfies the  DEC \eqref{dec}.  The proof of Theorem \ref{rigid2} above turns out to be a rather mild variation of the proof of \cite[Theorem 3.1]{Grigid}.

\proof[Proof of Theorem \ref{rigid2}]  As observed in Section \ref{prelim}, since $\S$ is weakly outermost, it is stable, $\l_1(L) \ge 0$.  Then, since $\S$ does not admit a metric of positive scalar curvature, Lemma \ref{scalar} applied to $L$, with $P = (\mu + J(\nu)) + \frac12 |\chi|^2$, implies that $\l_1 = 0$.  Hence, there exists a neighborhood $W = (-t_0,t_0) \times \S$ of $\S$ with the properties specified in Lemma \ref{foliation}. 
In particular, for each $t \in (-t_0,t_0)$, the outward null expansion $\th = \th(t)$ of $\S_t$ is constant.

A computation  similar to that leading to (\ref{thder}) (but where we can no longer
assume $\th$ vanishes)
shows that the null expansion function $\th = \th(t)$ of the foliation obeys the evolution equation (see \cite{AEM}),
\begin{align}\label{evolve}
\frac{d\th}{dt}  =  
 -\triangle \phi + 2\<X,\D\phi\> +  \left(Q   - \frac12 \th^2  + \th\tau
+{\rm div}\, X - |X|^2 \right)\phi \, ,
\end{align}
where it is to be understood that, for each $t$, the above terms  live on $\S_t$,  
e.g., $\triangle = \triangle_t$ is the Laplacian on $\S_t$, 
$Q =Q_t$ is the quantity \eqref{Q} now defined on $\S_t$, etc.  Also, in the above,  $\tau$ is the mean curvature of $V$.

The assumption that $\S$ is weakly outermost, together with the constancy of $\th(t)$, implies that $\th(t) \ge 0$ for all $t \in [0,t_0)$.   Fixing  $\e \in (0, t_0)$, we will now show that $\th(t) = 0$ for all $t \in [0,\e)$.
To this end, we re-express \eqref{evolve}  as follows,
\beq\label{evolve2}
\frac{d\th}{dt} -\tau\phi \,\th = L_t(\phi)  \,,
\eeq
where
\beq\label{newop}
L_t(\phi) =  -\triangle \phi + 2\<X,\D\phi\> +  \left(Q  - \frac12 \th^2  +{\rm div}\, X - |X|^2 \right)\phi \,.
\eeq
On $[0,\e] \times \S$, fix a constant  $c$ such that $\tau \phi \le c$.  Then \eqref{evolve2} and the nonnegativity of $\th$ imply,
\beq\label{opineq}
L_t(\phi) \ge \frac{d\th}{dt} - c \th = e^{ct} \frac{d}{dt}F(t)  \,, \quad \text{for all $t\in [0,\e)$}  \,,
\eeq 
where $F(t) = e^{-ct} \th(t)$.  We have that $F(0) = 0$ and $F(t) \ge 0$ on  $[0,\e)$.  To show that $F(t) = 0$ on 
$[0,\e)$, it is sufficient to show that $F'(t) \le 0$ for all $t \in [0,\e)$.  

Suppose there exists $t \in [0,\e)$, such that $F'(t) > 0$. Then \eqref{opineq} implies that $L_t(\phi) > 0$, and so,  by Lemma \ref{prin}, $\l_1(L_t) > 0$.  Applying Lemma \ref{scalar} to the operator $L_t$, where, in this case, $P=P_t = (\mu + J(\nu)) +\frac12 |\chi|^2 + \frac12 \th^2 \ge 0$, $\S_t \approx \S$, carries a metric of positive scalar curvature, contrary to assumption. 

Thus, $F(t) = 0$, and hence, $\th(t) = 0$ for all $t \in [0,\e)$.  Since, by \eqref{evolve2},  $L_t(\phi) = \th'  - \t\phi\th= 0$, Lemma \ref{prin} implies  $\l_1(L_t) \ge 0$ for each $t \in [0,\e)$. Hence, by Lemma~(\ref{scalar}), we have that  for each $t \in [0,\e)$, $\chi_t  = 0$ and $\S_t$ is Ricci flat.\qed 

\medskip
Theorem \ref{rigid2} has the following immediate consequence.

\begin{thm}\label{posyam3}
Let $(V^n,h,K)$, $n \ge 3$, be an initial data set satisfying the DEC, $\mu \ge |J|$. 
If  $\S^{n-1}$ is an outermost  MOTS in $(V^n,h,K)$ then $\S$ admits a metric of positive scalar curvature. 
\end{thm}

\smallskip 
We remark in closing that Theorems \ref{rigid2} and \ref{posyam3} may be viewed as local results in the following sense.   Let $\S$ be a MOTS in an initial data set satisfying the DEC.  By definition, $\S$ is $2$-sided.  As such, $\S$ admits a neighborhood $U$, within which $\S$ is separating.  To apply Theorem \ref{rigid2} (resp., Theorem \ref{posyam3}) it is sufficient that $\S$ be weakly outermost (resp., outermost) in $U$.


\providecommand{\bysame}{\leavevmode\hbox to3em{\hrulefill}\thinspace}
\providecommand{\MR}{\relax\ifhmode\unskip\space\fi MR }
\providecommand{\MRhref}[2]{%
  \href{http://www.ams.org/mathscinet-getitem?mr=#1}{#2}
}
\providecommand{\href}[2]{#2}

\end{document}